\documentclass[a4paper,10pt,twoside]{cpc-hepnp}

\usepackage{multicol}
\usepackage{graphicx}
\usepackage{booktabs}
\usepackage{amssymb,bm,mathrsfs,bbm,amscd}
\usepackage[tbtags]{amsmath}
\usepackage{lastpage}
\usepackage{CJK}

\begin{document}
\begin{CJK*}{GBK}{song}

\title{Spatial Dependent Diffusion of Cosmic Rays and the Excess of Primary Electrons Derived from High Precision Measurements by AMS-02}

\author{%
      C. Jin $^{1,2}$
\quad Y. Q. Guo $^{2}$
\quad H. B. Hu $^{2}$
}
\maketitle

\address{%
$^1$ School of Physical Engineering , Zhengzhou University, Zhengzhou 450001, China\\
$^2$ Key Laboratory of Particle Astrophysics, Institute of High Energy Physics, Chinese Academy of Sciences, Beijing 100049, China\\
}

\begin{abstract}
The precise spectra of Cosmic Ray (CR) electrons and positrons have been published by the measurement of AMS-02. It is reasonable to regard the difference between the electron and positron spectra ( $\triangle \Phi= \Phi_{e^-}-\Phi_{e^+}$ ) as being dominated by primary electrons. The resulting electron spectrum shows no sign of spectral softening above 20 GeV, which is in contrast with the prediction of the standard model of CR propagation. In this work, we generalize the analytic one-dimensional two-halo model of diffusion to a three-dimensional realistic calculation by implementing spatial variant diffusion coefficients in the DRAGON package. As a result, we can reproduce the spectral hardening of protons observed by several experiments, and predict an excess of high energy primary electrons which agrees with the measurement reasonably well. Unlike the break spectrum obtained for protons, the model calculation predicts a smooth electron excess and thus slightly over-predicts the flux from tens of GeV to 100GeV. To understand this issue, further experimental and theoretical studies are necessary.
\end{abstract}

\begin{multicols}{2}

\section{Introduction}
With unprecedentedly high precision, the AMS02 collaboration has recently confirmed the excess of positron fraction previously observed by PAMELA \cite{pamela1, pamela2}. The data shows a discrepancy from the prediction of the standard model of cosmic ray (CR) propagation (Conventional Model or CM), where the number of events dramatically rises at energies $\sim 10-300 \ GeV$, which is concluded to be an excess of positrons. Proposed explanations include dark matter annihilations/decay \cite{DM1, DM2}, extra sources such as pulsars \cite{pulsar1, pulsar2}, production of secondary leptons occuring around acceleration sources \cite{pp1, pp2}, or interactions between high-energy CRs and background photons \cite{photons, young}. The precise measurement of AMS also gives us an opportunity to study the spectrum of primary electrons. Dark matter annihilation/decay and nearby pulsar acceleration would produce exccess components of positrons and electrons in equal quantities, but because the number produced in secondary production by cosmic ray interaction with gases is orders of magnitude smaller than the number of primary electrons, it is reasonable to treat the difference between the flux of electrons and positrons as arising from the flux of primary electrons only ( $\triangle \Phi = \Phi_{e^-}- \Phi_{e^+}$ ). As is pointed out by \cite{expriel}, the spectrum of primary electrons obtained in this way exhibits a roughly constant spectral index for energies above $\sim$ 20 GeV, while the standard model predicts a continuously softening spectrum. This anomaly suggests an excess of primary electrons. In order to give a good fit to the data, a two-break injection spectrum was found inevitable with the allowed parameter space in the CM \cite{exelyq, expriel, exel}.
\par
On the other hand, a remarkable spectral hardening of primary CR nuclei and protons at several hundred GeV has been revealed by ATIC \cite{aticsol}, CREAM \cite{creamsol}, and PAMELA \cite{pamelasol}. Possible explanations of this spectral hardening include the spatial dependent diffusion of cosmic rays \cite{thm, dragondif1}, dual acceleration mechanism at the shock of SNRs \cite{accSNR}, interaction of cosmic rays with shock wave \cite{shock}, additional contribution from nearby supernova remnants (SNRs) \cite{nearSNR}, the effect of re-acceleration of cosmic rays by weak shocks associated with old supernova remnants in the Galaxy \cite{old} and so on.
\par
As assumed in the CM, CRs undergo a uniform diffusion in the galaxy, which provides a good but possibly oversimplified approximation of CR propagation. The CR diffusion is due to their random scattering by hydromagnetic waves and depends on magnetic field irregularities. On the Galactic disk, the irregularities are caused by SNR explosions, while in the diffusive halo, the irregularities are generated by CRs themselves in the absence of SNRs. It has been found that the spectra of the turbulences in the far outer galaxy and in the halo should be flatter than that in the local or inner galaxy  \cite{evidenceofspacechange}, which implies strong spatial changes of CR diffusion properties. In this work, we study the impact of latitudinal variation of CR diffusion properties on the spectrum of primary electrons, and generalize the Two-Halo Model (THM) \cite{thm} to a three-dimensional realistic calculation by implementing spatial variant diffusion coefficients in DRAGON package \cite{gamma}. For comparison, we also perform a CM calculation with DRAGON, which assumes a uniform diffusion of CRs in space. First, we reproduce the spectral hardening of primary protons to determine the relevant parameters, and then we attempt to study the excess of electrons with this scenario.

\section{Modeling and parameter settings}
It has been long considered that SNRs are the origin of primary Galactic Cosmic Rays, and the diffusive shock acceleration is regard as the main acceleration mechanism. CRs are accelerated at SNRs and then diffuse in the Galaxy, suffering from fragmentation and energy losses in the interstellar medium (ISM) and interstellar radiation field (ISRF) and magnetic field, decay and possible reacceleration or convection. Considering those processes, the CRs propagation equation can be written as
\begin{eqnarray}\label{eq:crdiff}
\frac{\partial \psi(\vec r, p, t)}{\partial t} && =Q(\vec r,p,t)+ \nabla \cdot (D_{xx} \nabla \psi- V_{c} \psi) \nonumber \\ [1mm]
&& + \frac{\partial}{\partial p}p^2D_{pp}\frac{\partial}{\partial p}\frac{1}{p^2}\psi -\frac{\partial}{\partial p}[\dot p \psi- \frac{p}{3}(\nabla \cdot V_c \psi)] \nonumber \\ [1mm]
&& -\frac{\psi}{\tau_f}-\frac{\psi}{\tau_r}
\end{eqnarray}
where $\psi(\vec r, p, t)$ is the density of CR particles per unit momentum p at position $\vec r$, $Q(\vec r, p, t)$ is the source distribution, $D_{xx}$ is the spatial diffusion coefficient, $V_c$ is the convection velocity, $D_{pp}$ is the diffusive reacceleration coefficient in momentum space, $\dot p \equiv \frac{dp}{dt}$ is momentum loss rate, and $\tau _f$ and $\tau_r$ are the characteristic time scales for fragmentation and radioactive decay respectively. In the CM, CR diffusion is assumed to be uniform in space and only energy-dependent , and the diffusion coefficient is parametrized as $D_{xx}=\beta D_0 (\rho/\rho_0)^\delta$, a function where $\rho$ is the rigidity and $\delta$ reflects the property of the ISM turbulence. The reacceleration can be described by the diffusion in momentum space and the momentum diffusion coefficient $D_{pp}$ is coupled with the spatial diffusion coefficient $D_{xx}$ as \cite{acc}
\begin{equation}\label{acc}
D_{pp}D_{xx}=\frac{4p^2v_A^{2}}{3\delta(4-\delta^2)(4-\delta)\omega}
\end{equation}
where $v_A$ is the Alfven speed, and $\omega$ is the ratio of magnetohydrodynamic wave energy density to magnetic field energy density, which can be fixed to 1. In this work, we consider a space-dependent diffusion process, and take the parameter $\delta$ of the diffusion coefficient as a variation along the z axis with $\delta(z)$.
\par
To illustrate the effect of spatial change on CR diffusion properties, we first consider a simplified situation. If we only consider the diffusion process during CRs propagating in the galaxy and neglect other processes, as is described in \cite{thm}, it can be solved using 1D analytical calculations. The proposed THM considers the latitudinal changes of CR diffusion properties , and it divides the diffusive halo with a typical half-thickness $L\sim 5 \ kpc$ into two regions, where the inner halo is influenced mainly by SNRs, with a half-thickness $\xi L \ (\xi \sim 0.1)$, and the outer halo dominates the rest of the wide region driven by CRs themselves. The diffusion coefficient is expressed as
\begin{equation}\label{eq:dirichlet}
D(z,\rho) = \begin{cases}
D_0 \beta (\frac{\rho}{\rho_0})^{\delta}, & \text{for } \lvert z \rvert < \xi L ( \ \text{inner halo} \ ) \\
D_0 \beta (\frac{\rho}{\rho_0})^{\delta+\triangle}, & \text{for } \lvert z \rvert > \xi L ( \ \text{outer halo} \ )
\end{cases}
\end{equation}
where $\rho$ is rigidity and $D_0$ specifies its normalization at the reference rigidity $\rho_0$. This model behaves like a reservoir, where CRs leak out rapidly into the outer halo and can re-enter the inner halo. As the result, the spectrum of CR primary species is read as
\begin{equation}\label{sol}
N_0 \equiv N(z=0) \sim \frac{L}{D_0}\{ \xi \rho^{-\nu-\delta} + (1-\xi) \rho^{-\nu-\delta-\triangle} \}
\end{equation}
which obviously indicates two components of the spectrum. We can derive from Eq. (\ref{sol}) that the inner halo dominates the high-energy part with the factor $\xi$, and the outer halo mainly contributes CRs at low energies with the factor $(1-\xi)$. Because of the relatively larger diffusion coefficient in the outer halo, CRs possess a higher escape velocity, which causes fast leakage, and only the low-energy CRs re-enter the Galactic disk to the observer's position. Meanwhile, the inner halo with its smaller diffusion coefficient can detain more CRs at high energies. As a consequence, we can derive the spectral hardening of primary protons from the spatial change of CR diffusion properties.
\par
In this work, we use the released DRAGON code to solve the CR propagation equation described in Eq. (\ref{eq:crdiff}). DRAGON allows us to perform a numerical calculation with a space-dependent coefficient. Without loss of generality, we take the diffusion coefficient index $\delta(z)$ as the form
\begin{equation}\label{delta2}
\delta(z) = \begin{cases}
\delta_0+\triangle( \frac{z}{\xi L})^n , & \text{for } \lvert z \rvert < \xi L ( \ \text{inner halo} \ ) \\
\delta_0+\triangle, & \text{for } \lvert z \rvert > \xi L ( \ \text{outer halo} \ )
\end{cases}
\end{equation}
where we use a power-law function to describe the gradual transition from inner halo to outer halo and the index n is fixed at 5, representing the extent of how sharply $\delta_0$ changes to $\delta_0+\triangle$. We extend the size of the inner halo to $\xi L$ with $\xi \sim 0.16$ to contain the highest number of SNRs. For comparison, we carry out calculations for both THM and CM using DRAGON, the CM being set with a uniform diffusion coefficient in space. Detailed information for the parameters of the CR propagation equation is shown in Table 1.
\begin{center}
\tabcaption{ \label{tab1}  Some parameters of the propagation equation}
\footnotesize
\begin{tabular*}{80mm}{c@{\extracolsep{\fill}}ccc}
\toprule Parameters & THM   & CM\\
\hline
$D_0 \  (cm^2 \ s^{-1})$ & $5.5\times10^{28}$ & $5.5\times10^{28}$ \\
$v_A \ (km/s)$ & 25 & 25 \\
$V_c \ (km/s)$ & 0 & 0 \\
$\delta$ & 0.19 & 0.5 \\
$\triangle$ & 0.39 & \\
\bottomrule
\end{tabular*}
\end{center}
\par
The injection spectrum of primary CRs at the source region is taken as a broken power-law form
\begin{equation}\label{inj}
Q_i(E_k,r,z)=f_s(r,z)q_0^i \times \begin{cases}
(\frac{\rho}{\rho_{br}})^{-\alpha_0}\exp(-\frac{\rho}{\rho_{cut}}) & \text{if } (\rho < \rho_{br}) \\
(\frac{\rho}{\rho_{br}})^{-\alpha_1}\exp(-\frac{\rho}{\rho_{cut}}) & \text{if } (\rho > \rho_{br})
\end{cases}
\end{equation}
where $\rho_{br}$ is the break position of rigidity, and $\rho_{cut}$ represents cutoff rigidity.
The normalization condition $f_s(r_{\odot},z_{\odot})=1$ is imposed, where $r_{\odot}=8.5 \ kpc$ is the distance from the Sun to the Galactic center. The spatial distribution of SNRs $f_s(r,z)$ is modeled as in \cite{snrmodel}. Detailed information for the injection spectrum of primary protons and electrons is listed in Table 2.
\begin{center}
\tabcaption{ \label{tab1}  Injection spectrum of primary CRs}
\footnotesize
\begin{tabular*}{80mm}{c@{\extracolsep{\fill}}ccc}
\toprule Parameters & Proton   & Electron\\
\hline
$\rho_{br} (GV)$ & 9.0 & 5.7 \\
$\alpha_0$ & 2.1 & 1.7 \\
$\alpha_1$ & 2.31 & 2.7 \\
$\rho_{cut} (GV)$ & 2.5e6 & 20000 \\
\bottomrule
\end{tabular*}
\end{center}
\par
Following the parameters fixed above, we obtain the spectrum of primary protons. The result is shown in Fig.~\ref{fig1}.
\begin{center}
\includegraphics[width=8cm]{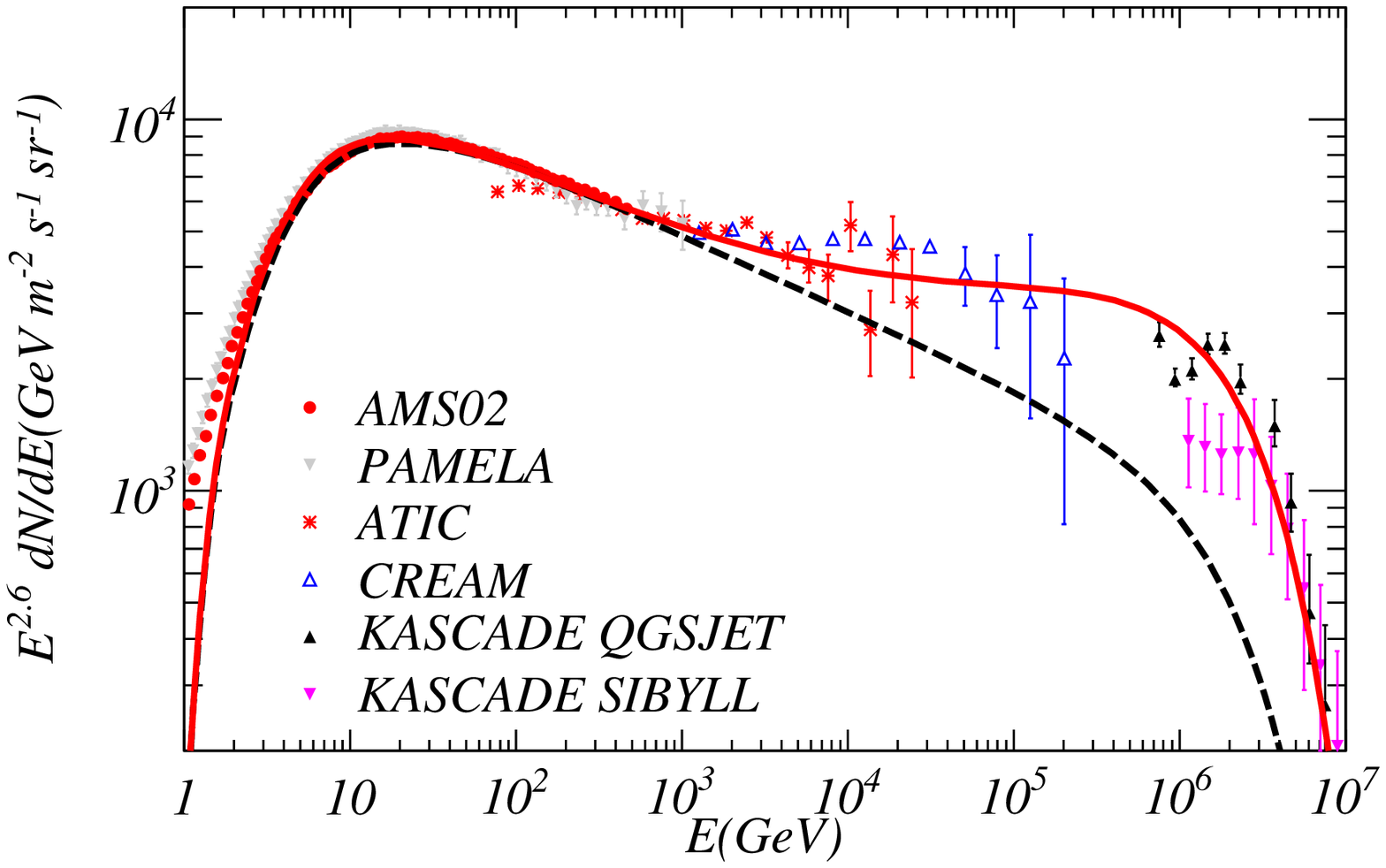}
\figcaption{\fontsize{9}{10}\selectfont \label{fig1}   The proton spectra from measurements and the two model calculations. The experiment data for the proton come from: AMS02 \cite{AMSdata}, ATIC \cite{ATICdata}, PAMELA \cite{pamelasol}, CREAM \cite{CREAMdata}, KASCADE \cite{KASCADEdata} }
\end{center}
\par
The red solid line represents the THM prediction, and the black dashed line represents the CM calculation. In both calculations, the parameters are tuned to fit well with the AMS data from $10-10^3 \ GeV$. At energies above $\sim 10^3 \ GeV$, the THM exhibits a pronounced spectral upturn above $\sim 10^3 \ GeV$, and is in good agreement with the data within uncertainties. It should be noted that although spectral hardening above $\sim 10^3 \ GeV$ can be recovered, the sharp transition of spectral slope at $\sim 200 \ GeV$ found by PAMELA \cite{pamelasol} cannot be reproduced.
As a matter of fact, when the energy is below $\sim 10 \ GeV$, the solar modulation comes to play a major role in the spectrum. The effect of solar modulation is taken care of by a so-called force-field approximation \cite{solarmod}, with the modulation potential being 550 MV.
\par
As described above, the effect of spatial change on CR diffusion properties is confirmed to produce the spectral hardening of primary protons by a realistic three-dimensional model. The observed data above $10^3 \ GeV$ support this assumption of a THM. In the next section, we will show the result for primary electrons with same set of diffusion parameters.

\section{Results of primary electrons}
Unlike primary protons, primary electrons behave in a more complicated way. Propagating electrons suffer severe energy loss by processes such as synchrotron radiation and inverse Compton scattering, which will rapidly steepen the spectrum of electrons with increasing energies. The power of energy loss is almost proportional to $E^2$, so higher energy electrons loss energy rapidly. TeV electrons lose most of their energy on a timescale of $\sim 10^5 \ yr$, so only those within a limit of several hundred parsecs can reach the solar system. Comparing with the power-law or upturn spectra for primary protons in Fig. 1, the primary electron spectra are heavily attenuated for both THM and CM. The corresponding spectra are shown in Fig. 2, together with the subtraction between electron and positron flux $(\triangle \Phi=\Phi_{e^-}-\Phi_{e^+})$ measured by AMS-02.
\begin{center}
\includegraphics[width=8cm]{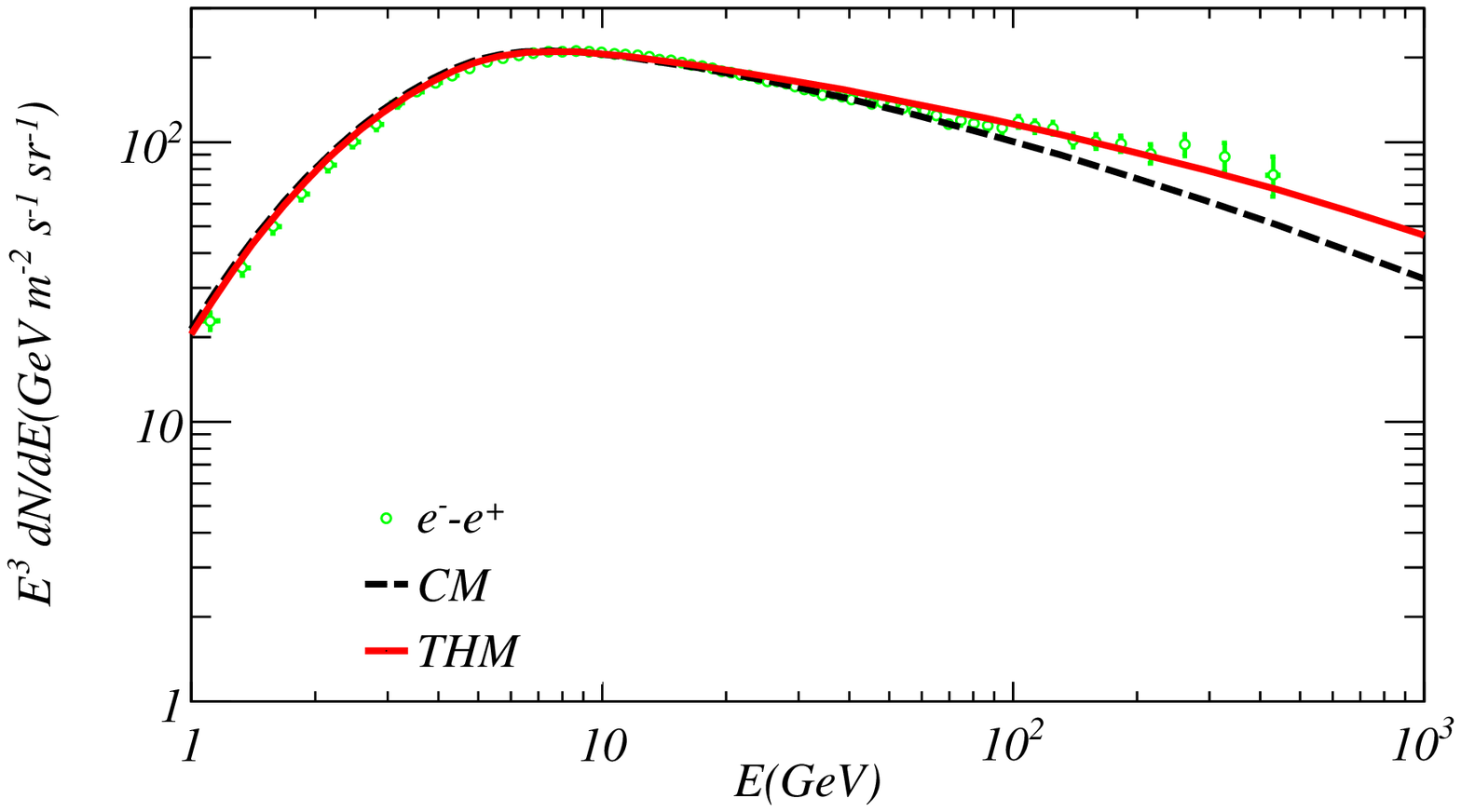}
\figcaption{\fontsize{9}{10}\selectfont \label{fig2}   The electron spectrum in the two models. $\phi_{e^+}$ and $\phi_{e^-}$ data are from AMS02 \cite{positron, positronpluseminus} }
\end{center}
In Fig. 2, the observed spectrum seems to be slightly upturned at $\sim 100 \ GeV$, but there is no sign of a spectral break in the THM (red solid line). The standard model (black dashed line) can reproduce the data well below $\sim 40 \ GeV$, but becomes insufficient as energy increases. The spectrum of primary electrons predicted by the THM reveals an integral hardening above $\sim10 \ GeV$, which makes it successfully recover the data within error bars. However, it can be seen that the THM overestimates the flux of primary electrons slightly with respect to the data $\triangle \phi$ between 30 GeV and 100 GeV.
\par
Due to rapid energy losses, primary electrons are strongly affected by the cooling time, while the influence of cooling time on primary protons can be neglected above 10 GeV. The cooling timescale of electrons can be estimated by $\tau_c \sim 17 \ Myr \ (\epsilon_e / 10 \ GeV)^{-1}$ \cite{expriel}, and electrons can diffuse a distance of $R=(2D \tau_c)^{1/2}$ in this time, where D is the diffusion coefficient. At 10 GeV, the distances can be calculated as $\sim 3.1 \ kpc$ in the CM, $\sim 2.7 \ kpc$ in the inner halo and $\sim 3.2 \ kpc$ in the outer halo in the THM. It can be seen that the effect of rapid energy losses narrows the effective halo size with respect to the fixed half-thickness 5 kpc of the halo. The effective halo size will continuously shrink with increased energy, so the contribution of the inner halo to the total spectrum will increase. As described in the former simplified 1D analysis, the inner halo dominates the high energy components with a factor $\xi$ coupled with the contribution of the inner halo in the steady state. When the factor $\xi$ rises, relatively more high-energy CRs remain in the inner halo and harden the final spectrum. In this scenario, the primary electrons predicted by the THM will exhibit an integral hardening spectrum above $\sim10 \ GeV$.

\section{Discussion and conclusion}
In this work, we focus on the latitudinal changes of CR diffusion properties and reproduce the spectral hardening of primary protons above $\sim 10^3 \ GeV$ in this scenario. However, the obtained spectrum of primary electrons exhibits an integral hardening with no visible upturned break, which gives a good fit with the data above $\sim 100 \ GeV$ and overestimates a little between 30 GeV and 100 GeV. It should be noted that, although the THM produces a non-break spectrum similar to the CM prediction, the additional flux obtained with the THM cannot be reached by simply adjusting the global parameters used in the CM. The rapid energy losses of CR electrons are responsible for the difference between the spectrum of primary protons and electrons.
\par
In addition, it has been pointed out that a continuous source distribution model is not valid for electrons in the energy region above 100 GeV \cite{discrete}, while we still consider a continuous distribution in numerical calculation. The spectrum of electrons at high energies should depend on the age and distance of a few nearby sources, and some nearby discrete sources have been considered to explain the spectrum of electrons \cite{discrete1, discrete2}. In the meanwhile, the space-dependent effect of CR diffusion properties has been used to explain the excess spectra widely observed for secondaries \cite{fresh}, the longitude profile of the diffuse $\gamma - ray$ emission \cite{dragondif1, gamma}, and the significant fraction of the IceCube neutrino flux \cite{dragondif1}. It is worth taking both the effect of spatial change on CR propagation and nearby discrete sources into account in the future.

\acknowledgments{This work is supported by the Natural Sciences Foundation of China (11135010).}

\vspace{10mm}
\end{multicols}

\vspace{-1mm}
\centerline{\rule{80mm}{0.1pt}}
\vspace{2mm}

\begin{multicols}{2}

\end{multicols}

\clearpage

\end{CJK*}
\end{document}